\documentclass
[aps,pre,showpacs]{revtex4}
\usepackage{graphicx}
\begin{document}
\title{Dark solitons in ferromagnetic chains with first- and second-neighbor interactions}

\author{M. T. Primatarowa and R. S. Kamburova}

\address{Georgi Nadjakov Institute of Solid State Physics, Bulgarian Academy of Sciences,
1784 Sofia, Bulgaria}

\begin{abstract}
We study the ferromagnetic spin chain with both first- and second-neighbor
interactions. We obtained the condition for the appearance and stability of
bright and dark solitons for arbitrary wave number inside the Brillouin zone.
The influence of the second-neighbor interaction and the anisotropy on the
soliton properties is considered. The scattering of dark solitons from point
defects in the discrete spin chain is investigated numerically.
\end{abstract}


\maketitle

\section{Introduction}
Considerable attention has been devoted to the study of solitary waves in
solids. A large amount of theoretical and experimental research has been
dedicated to one-dimensional magnetic systems [1]. Models corresponding to real
quasi-one-dimensional magnets are broadly investigated as first they were
considered as somewhat simpler than the really interesting three-dimensional
systems but then turned out to be interesting in their own right. There was the
availability of rear magnetic compounds which because of extremely small
coupling between neighboring chains, could be considered as reasonable
realizations of magnetically one-dimensional materials. Soliton solutions for
classical sine-Gordon chains [2] and Heisenberg chains with various
anisotropies [3-6] were obtained and analyzed. In recent years there is a
renewed interest to the topic due to their application. Spin-wave dark solitons
were predicted and experimentally generated [7,8]. Solitons in magnetic thin
films [9] and in ferromagnets with biquadratic exchange [10] were investigated.

Bright and dark solitons as exact solutions of the nonlinear Schr\"{o}dinger
(NLS) equation have been studied intensively [3,4,11]. But, the models
describing the physical effects in solids are mainly discrete and the question
arises how the discreteness modify the properties of the nonlinear excitations
[12,13]. Another interesting topic of research with practical importance is the
interaction of solitons with impurities [14-18]. The influence of the
interaction with second-neighbors [19] and third-neighbors [20] on the bright
soliton dynamics has been also considered.

In the present paper we study the condition for the appearance of dark solitons
in an anisotropic ferromagnetic chain with first- and second-neighbor
interactions. The propagation for arbitrary wave number is considered and the
scattering on point defects is investigated.

\section{Hamiltonian of the system}
We consider a ferromagnetic Heisenberg chain of $N$ spins with magnitude $S$
 described by the
following Hamiltonian:
\begin{eqnarray}
H &=& -\frac{J_1}{2}\sum_{n}(S_{n}^{+} S_{n+1}^{-}+S_{n}^{-} S_{n+1}^{+})
-\tilde{J_1}\sum_{n}S_{n}^{z} S_{n+1}^{z}\nonumber\\
&-&\frac{J_2}{2}\sum_{n}(S_{n}^{+} S_{n+2}^{-}+S_{n}^{-} S_{n+2}^{+})
-\tilde{J_2}\sum_{n}S_{n}^{z} S_{n+2}^{z} - A\sum_{n}(S_{n}^{z})^2 \, ,
\end{eqnarray}
where both nearest-neighbor and next-nearest-neighbor exchange interactions are
taken into account. The spin operators $S_{n}^{\pm}$, $S_{n}^{z}$ obey the
commutation relations
\begin{equation}
[S_{i}^{\pm},S_{j}^{z}]=\mp S_{i}^{\pm} \delta_{ij},\quad
[S_{i}^{+},S_{j}^{-}]=2S_{i}^{z} \delta_{ij}\, .
\end{equation}
$J_i>0$ and $\tilde{J_i}>0$  $(i=1,2)$ are the exchange integrals and $A$ is
the on-site anisotropy constant which can be positive (easy axis) or negative
(easy plane). If $\tilde{J_i}\neq J_i$ the inter-site anisotropy is included.
For $J_i=\tilde{J_i}$ and $A=0$ the model is isotropic.

The equations of motion for the operators $S_n^{+}$ yield $(\hbar = 1$):
\begin{eqnarray}
i\frac{\partial S_n^{+}}{\partial t} & = &- A S_{n}^{+}
 -S_{n}^{z}[J_1(S_{n+1}^{+}+S_{n-1}^{+})+J_2(S_{n+2}^{+}+S_{n-2}^{+})]\nonumber\\
&+&[\tilde{J_1}(S_{n+1}^{z}+S_{n-1}^{z})+\tilde{J_2}(S_{n+2}^{z}+S_{n-2}^{z})]
S_{n}^{+} + 2A S_{n}^{z}S_{n}^{+} \, .
\end{eqnarray}

In the quasiclassical approximation, where the components of the spin operators
$S_n$ are complex amplitudes, $\alpha_n=S_{n}^{+}/S$,
$\alpha_n^{\ast}=S_{n}^{-}/S$ and $S_{n}^{z}/S=\sqrt{1-|\alpha_{n}|^{2}}$, we
have
\begin{eqnarray}
i\frac{\partial \alpha_n}{\partial t} &=& -A \alpha_{n}
 - [J_1 S (\alpha_{n+1}+\alpha_{n-1}) + J_2 S (\alpha_{n+2}+\alpha_{n-2})]
 \sqrt{1-|\alpha_{n}|^{2}}\nonumber\\
&+& \alpha_{n}[\tilde{J_1}S
\left(\sqrt{1-|\alpha_{n+1}|^{2}}+\sqrt{1-|\alpha_{n-1}|^{2}}\right)
+\tilde{J_2}S
\left(\sqrt{1-|\alpha_{n+2}|^{2}}+\sqrt{1-|\alpha_{n-2}|^{2}}\right)]\nonumber\\
 &+& 2AS\alpha_{n}\sqrt{1-|\alpha_{n}|^{2}} \, .
\end{eqnarray}

The set of differential equations (4) describes our system. In the continuum
limit (wide excitations), it transforms into a perturbed NLS equation, where
the perturbing terms are not only due to the discreteness but also to the
complicated nonlinear interactions. For $J_2=\tilde{J_1}=\tilde{J_2}=A=0$ and
wide excitations, (4) turns to the discrete Ablowitz-Ladik equation which is
also completely integrable and has soliton solutions. So, for narrow
excitations it is necessary to investigate the whole set (4). Equations similar
to (4) has been derived for exciton solitons in molecular crystals [21].

\section{Soliton solutions}

We shall look for solutions in the form of amplitude-modulated waves
\begin{equation}
\alpha_{n}(t) = \varphi_n(t)e^{\textstyle i(kn-\omega t)} \,,
\end{equation}
where $k$ and $\omega$ are the wave number and the frequency of the carrier
wave (the lattice constant equals unity) and the envelope $\varphi_n(t)$ is a
slowly varying function of the position and time. In the continuum limit, when
the soliton width is much larger than the lattice spacing ($L\gg1$), equation
(4) transforms into the following NLS equation for the envelope:

\begin{equation}
i\left(\frac{\partial\varphi}{\partial \tau}+v_g
\frac{\partial\varphi}{\partial x}\right) = (\varepsilon-\Omega)\varphi - b_k
\frac{\partial^2\varphi}{\partial x^2} + g_k |\varphi|^{2}\varphi \, ,
\end{equation}
where
\begin{eqnarray}
\tau &=&tS\, ,\quad \Omega=\omega/S \, ,\quad g_k=J_1 \cos k - \tilde{J_1}+J_2
\cos 2k - \tilde{J_2}-A\, ,\quad \varepsilon = -A/S-2g_k\, ,\quad \nonumber\\
v_g&=&2(J_1\sin k+2 J_2\sin 2k)\, ,\quad b_k=J_1 \cos k+4J_2\cos 2k \, .
\end{eqnarray}
$v_g$ is the group velocity and $b_k$ describes the group-velocity dispersion
of the carrier waves. Depending on the sign of the expression $b_k g_k$
equation (6) possesses soliton solutions of different types. For negative
values bright solitons exist while for positive values dark solitons are
possible.

Let us consider the dependence of the coefficients $b_k$ and $g_k$ on $k$
(Fig.~1 and 2). The group-velocity dispersion $b_k$ depends only on $J_1$ and
$J_2$ and is not influenced by the anisotropy of the system. Without second
neighbor interaction ($J_2=0$) $b_k$ changes sign at the middle ($k=\pi/2$) of
the Brillouin zone. The inclusion of weak second-neighbor interaction
($J_2<1/4$) has influence only in the moving of this point to smaller
$k$-values. Strong second-neighbor interactions ($J_2>1/4$) lead to a second
point where the function $b_k$ changes again sign (Fig.~1). The nonlinear
coefficient $g_k$ depends on the anisotropy of the system. For the isotropic
case without second neighbor interaction ($J_1=\tilde{J_1}$,
$J_2=\tilde{J_2}=A=0$) $g_k$ is negative in the whole Brillouin zone. So in
this case equation (6) has bright-soliton solutions for $0<k<\pi/2$ and
dark-soliton solutions exist for $\pi/2<k<\pi$. The inclusion of positive
on-site anisotropy ($A>0$) preserves this behavior. The inclusion of negative
on-site anisotropy ($A<0$) leads to the appearance of a wave number $k_c$ so
that $g_k$ is positive for $0<k<k_c$ and negative for $k_c<k<\pi$ [(Fig.~2(a)].
Then the Brillouin zone is divided in three regions where the solution alters
from dark- to bright- and again to dark-soliton. If $k_c=\pi/2$ then the
solution of equation (6) is of the dark-soliton type in the whole Brillouin
zone. The inclusion of weak second-neighbor interactions has influence only on
the value of $k_c$ and the size of the regions where the different solutions
exist. Strong second-neighbor interactions with $J_2>1/4$ will lead to the
appearance of the additional region near the band edge $\pi$ where the function
$b_k$ becomes positive (Fig.~1) and consequently the solution is of the
bright-soliton type. Fig.~2(b) shows the function $g_k$  for the case of
inter-site anisotropy ($\tilde{J_1}=\tilde{J_2}=A=0$). When the second-neighbor
interactions are not taken into account the solution of equation (6) is of the
dark-soliton type in the whole Brillouin zone. The inclusion of second-neighbor
interactions will lead again to three or four (if $J_2>1/4$) regions where
different soliton solutions exist.

\begin{figure}
\resizebox{3in}{!} {\includegraphics{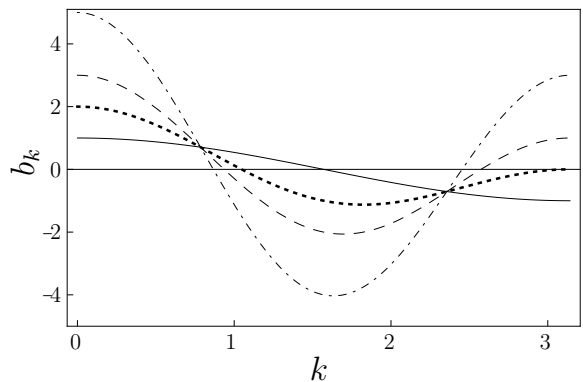}}
 \caption{\label{fig.1} Dependence of the group-velocity dispersion $b_k$ on
 the second-neighbor interaction. $J_1=1$; $J_2=0$ (solid line),
 $J_2=0.25$ (dotted line), $J_2=0.5$ (dashed line), and $J_2=1$ (dashed-dotted
 line). }
\end{figure}

\begin{figure}
\resizebox{3in}{!} {\includegraphics{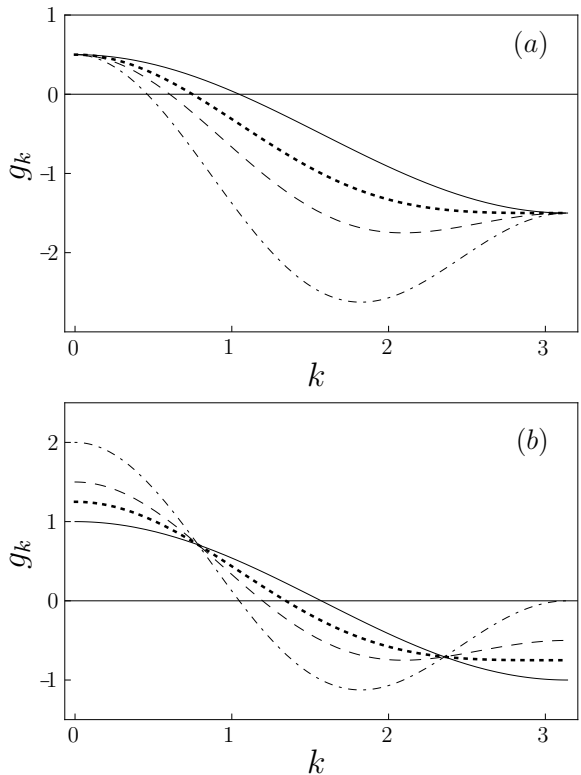}}
 \caption{\label{fig.2} Nonlinear coefficient $g_k$ for (a) on-site anisotropy
 ($J_i=\tilde{J_i}$, $A=-0.5$) and
(b) inter-site anisotropy ($\tilde{J_i}=A=0$); $i=1,2$. $J_2=0$ (solid lines),
$J_2=0.25$ (dotted lines), $J_2=0.5$ (dashed lines), and $J_2=1$ (dashed-dotted
lines). }
\end{figure}

In what follows, we shall consider dark-soliton solutions which appear for
\begin{equation}
b_k g_k >0
\end{equation}
and have nonvanishing boundary conditions, $|\varphi(x)|^{2}\rightarrow $ const
at $x\rightarrow\pm\infty$. Note that the condition (8) depends not only on the
anisotropy constants but also on the wave number $k$. The dark soliton has the
form
\begin{equation}
\varphi(x,\tau) = \varphi_0{\rm tanh} \frac{x-v\tau}{L}+iB\, ,
\end{equation}
where $L$ and $v$ are its width and velocity. $B$ determines the value of the
amplitude at the center, i.e the minimum intensity of the soliton
$|\varphi(0)|^{2}=B^2$, while the maximum intensity is
$|\varphi(\pm\infty)|^{2}=\varphi_0^2+B^2$. As independent parameters of the
soliton we can choose $L$, $k$ and $B$. Then $\varphi_0$, $\Omega$ and $v$ are
given by the relations:

\begin{equation}
\varphi_0^2=\frac{2 b_k}{g_k L^2}\, ,\quad \Omega = \varepsilon
+g_k(\varphi_0^2+B^{2})\, ,\quad v = v_g+v_d\, ,\quad
 v_d\equiv B \sqrt{2g_k b_k}\,.
\end{equation}
The dark soliton can be considered as an excitation of the background (carrier)
wave. Its velocity consists of the group velocity $v_g$ and a contribution
$v_d$ which depends on $B$. For fixed ($\varphi_0^2+B^2$) and $k$ solitons
moving faster have smaller amplitude $\varphi_0$.

The bright soliton solution which exists for $b_k g_k <0$ and
$|\varphi(x)|^{2}\rightarrow 0$ at $x\rightarrow\pm\infty$ has the form
\begin{equation}
\varphi(x,\tau) = \varphi_0{\rm sech} \frac{x-v\tau}{L}
\end{equation}
with
\begin{equation}
\varphi_0^2=-\frac{2 b_k}{g_k L^2}\, ,\quad \Omega = \varepsilon
-\frac{b_k}{L^2}\, ,\quad v = v_g \,.
\end{equation}

We like to make some remarks about the influence of the next-nearest-neighbor
interaction on the form and velocity of the soliton solution. The dependence of
the velocity on $k$ is shown on Fig.~3. The group velocity $v_g$ is defined in
the whole Brillouin zone and coincides with the velocity of the bright solitons
and of the dark solitons for $B=0$ [Fig.~3(a)]. It is independent of the
anisotropy of the model. The small second neighbor interactions change only the
value of the velocity, while strong second neighbor interactions ($J_2>1/4$)
can change also its sign. The velocity of the dark soliton itself $v_d$ is
shown on Fig.~3(b). It is nonequal zero in the regions where the dark solitons
exist and its sign is determined by $B$ as given in (10).

\begin{figure}
\resizebox{3in}{!} {\includegraphics{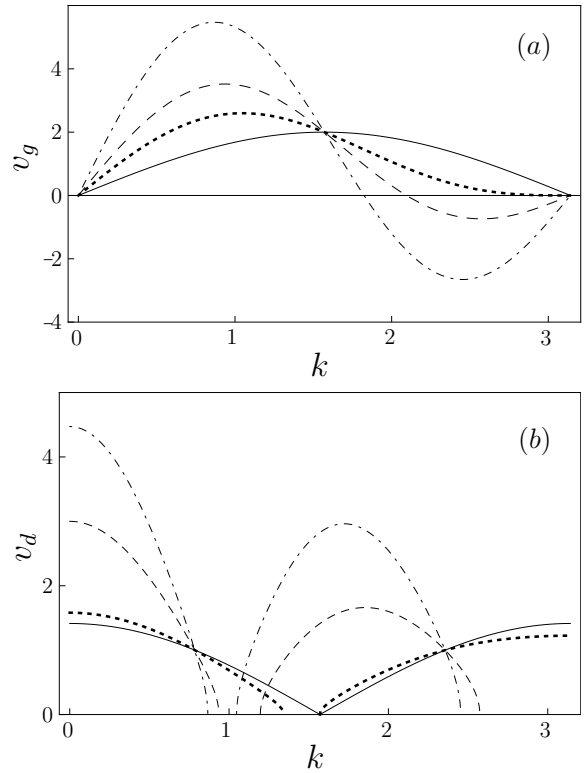}}
 \caption{\label{fig.3} k-dependence of the group velocity $v_g$ (a) and the
 dark-soliton velocity $v_d$ (b) for $\tilde{J_1}=\tilde{J_2}=A=0$. $J_2=0$ (solid
 lines), $J_2=0.25$ (dotted lines), $J_2=0.5$ (dashed lines), and $J_2=1$
 (dashed-dotted lines). }
\end{figure}

We performed numerical simulations based on the general discrete equation (4)
and we investigate mainly the dark soliton solution. As initial function we put
for $t=0$ the form
\begin{equation}
\alpha_{n}(0) = \left({\frac{\sqrt{2 b_k/g_k}}{L}\rm tanh}
\frac{n}{L}+iB\right)e^{\textstyle ikn}
\end{equation}
which is exact solution of the quasicontinuum equation, and observe the
evolution for different anisotropic cases. For wide solitons ($L\gg1$) this
solution remain stable and its form and velocity are preserved. When the
soliton width decreases the discreteness effects of equation (4) become
important and the parameters of the dark soliton change.

In what follows we have $L=10$ and $J_1=1$. Fig.~4 shows the propagation of the
dark solitons in the case $\tilde{J_i}=A=0$ for two $k$-values ($k_1=0.0157$
and $k_2=\pi-k_1$). When the second neighbor interaction is absent ($J_2=0$)
the solitons propagate with the same velocity and amplitude [Fig.~4(a),(a$'$)].
When the second neighbor interactions are present the solitons have different
amplitudes and velocities [Fig.~4(b),(b$'$) for $J_2=0.1$]. As can be seen from
the $k$-dependence of the velocity from Fig.~3 the value of the velocity for
the greater $k$ is smaller. The increase of the second neighbor interaction
leads to the turning of the solution for $k_2$ to a bright soliton type
[Fig.~4(c),(c$'$) for $J_2=0.3$].

Similar results have been obtained for the case with on-site anisotropy
$J_1=\tilde{J_1}$, $J_2=\tilde{J_2}$, $A\neq0$.

\begin{figure}
\resizebox{4in}{!} {\includegraphics{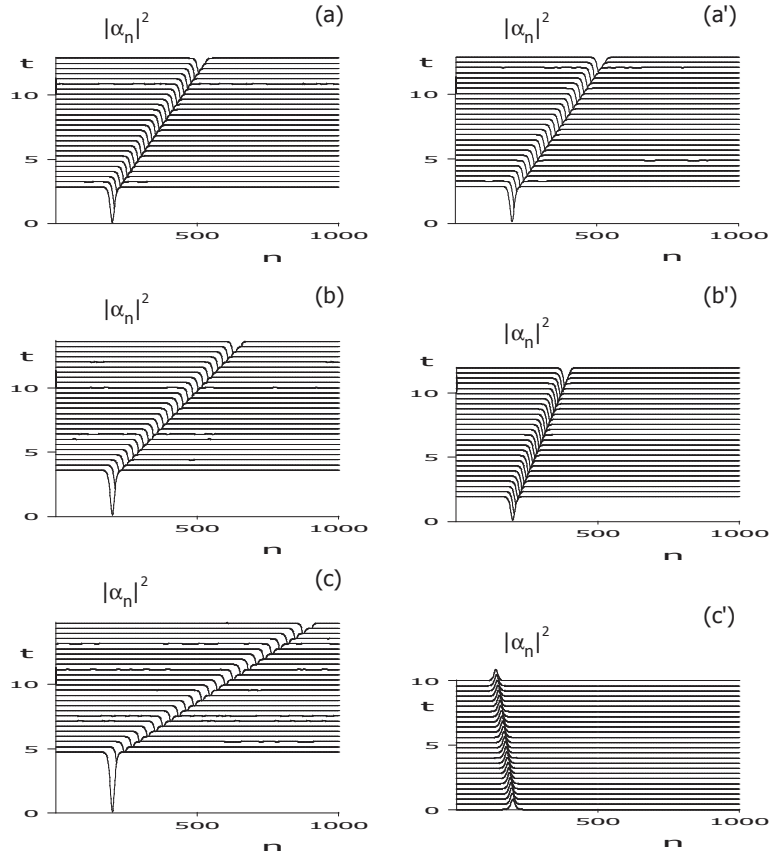}}
 \caption{\label{fig.4} Time evolution of dark solitons for inter-site
anisotropy $\tilde{J_1}=\tilde{J_2}=A=0$ and $k_1=0.0157$ [(a), (b), (c)],
$k_2=\pi-k_1$ [(a$'$), (b$'$), (c$'$)]. The time is in units of $10^3/J_1S$. }
\end{figure}

We like to point out that Eq.~(4) is invariant with respect to the
transformation $\alpha_{n} \rightarrow \alpha_{n}^* e^{\textstyle i\pi n}$ when
$J_2 \rightarrow -J_2$, $\tilde{J_1} \rightarrow -\tilde{J_1}$, $\tilde{J_2}
\rightarrow -\tilde{J_2}$, and $A \rightarrow -A$. So the results on
Fig.~4(b),(c) can be obtained also for $k_2$ provided $J_2$ has the values -0.1
and -0.3, respectively.

\section{Scattering on point defects}

We studied the interaction of the dark solitons with impurities in the chain
whose size is small compared to the soliton extent. Then the equation of motion
becomes
\begin{eqnarray}
 i\frac{\partial \alpha_n}{\partial t} &=& -A \alpha_{n}
 - [J_1 S (\alpha_{n+1}+\alpha_{n-1}) + J_2 S (\alpha_{n+2}+\alpha_{n-2})]
 \sqrt{1-|\alpha_{n}|^{2}}\nonumber\\
&+& \alpha_{n}[\tilde{J_1}S
\left(\sqrt{1-|\alpha_{n+1}|^{2}}+\sqrt{1-|\alpha_{n-1}|^{2}}\right)
+\tilde{J_2}S
\left(\sqrt{1-|\alpha_{n+2}|^{2}}+\sqrt{1-|\alpha_{n-2}|^{2}}\right)]\nonumber\\
 &+& 2AS\alpha_{n}\sqrt{1-|\alpha_{n}|^{2}} + d\delta_{n,n_0}\alpha_n\, .
\end{eqnarray}
The parameter $d$ describes the strength of the impurity at site $n_0$, which
can be attractive or repulsive. In our numerical simulations we investigate the
interaction of a propagating dark soliton of the form (13) with a defect placed
at the middle of the chain $n_0=500$. We have observed that in the case of
attraction the solitons pass through the defect for arbitrary initial velocity.
More interesting is the interaction with repulsive impurities.

In the case of repulsion the soliton can be transmitted or reflected by the
impurity. For a given initial velocity $v_0$ there is a critical value of the
defect $|d_c|$ below which the soliton is transmitted and above which it is
reflected. We consider dark solitons which velocities are determined only by
$k$ ($B=0$). Then, due to the symmetry properties of equations (4) and (14)
mentioned above, $d_c>0$ for small k-values and $d_c<0$ for $k$ near $\pi$.
Fig.~5 shows the scattering patterns for the case of inter-site anisotropy
($\tilde{J_1}=\tilde{J_2}=A=0$) and different wave numbers ($k_1=0.0157$ and
$k_2=\pi-k_1$). When the interaction with the second neighbors is absent
($J_2=0$) the scattering mechanisms for $k_1$ and $k_2$ are equal, as the
values for the initial velocity $v_0=0.031$ the values for $|d_c|$ are the same
and it holds: $0.0108<|d_c|<0.0109$. When the interaction with the second
neighbors is included ($J_2=0.1$) the initial velocities for the two cases
become different and the the values for $d_c$ are different, respectively. As
the initial velocity for $k_1$ ($v_0=0.044$) is greater than the initial
velocity for $k_2$ ($v_0=0.019$) the value for $|d_c|$ is higher. We observe
that the scattering mechanisms depends on the initial velocity and the strength
of the impurity. The anisotropy of the chain influences the amplitude of the
soliton in this case.
\begin{figure}
\resizebox{4in}{!} {\includegraphics{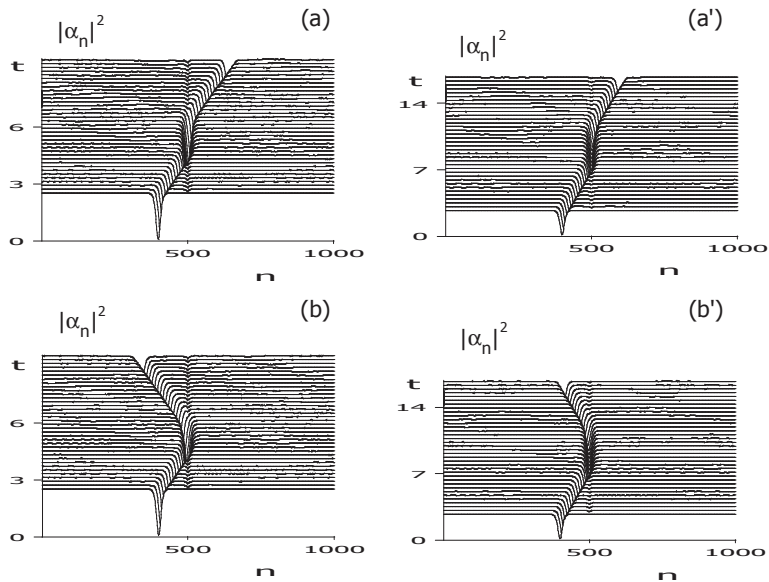}}
 \caption{\label{fig.5} Scattering of a dark soliton for inter-site anisotropy
 ($\tilde{J_1}=\tilde{J_2}=A=0$), $J_2=0.1$ and $k_1=0.0157$ [(a), (b)],
 $k_2=\pi-k_1$ [(a$'$), (b$'$)]. Transmission for $d=0.0152$ (a) and $d=-0.00652$ (a$'$).
 Reflection for $d=0.0153$ (b) and $d=-0.00653$ (b$'$). The time is in units of $10^3/J_1S$.}
\end{figure}

\section{Conclusion}

We have investigated the existence and stability of dark solitons in a discrete
ferromagnetic chain with both inter-site and on-site anisotropy. We have
considered the influence of the second-neighbor interactions on the soliton
dynamics in the whole Brillouin zone. The complicated dependence of the
dispersion and the nonlinear coefficients lead to regions in the Brillouin zone
where strong second-neighbor interactions can turn the type of the soliton
solution (from bright to dark or vice versa). The scattering of the dark
solitons on point defects is also investigated.

\begin{acknowledgments}
This work is supported in part by the National Science Fund of Bulgaria under
Grant No. DO 02-264.
\end{acknowledgments}


\begin{thebibliography}{99}

\bibitem {Mikeska} {H.-J. Mikeska and M. Steiner, Adv. Phys. {\bf 40},
191 (1991).}

\bibitem {Tjon} {J. Tjon and J. Wright, Phys. Rev. B {\bf 15}, 3470 (1977).}
\bibitem {Pushkarov} {D. I. Pushkarov and Kh. I. Pushkarov, Phys. Lett.
{\bf 61A}, 339 (1977).}

\bibitem {Huang} {G. Huang, Z.-P. Shi, X. Dai, and R. Tao, J. Phys.: Condens. Matter B
{\bf 2}, 8355 (1990).}

\bibitem {Wallis} {R. F. Wallis, D. L. Mills, and A. D. Boardman, Phys. Rev. B {\bf
52}, R3828 (1995).}

\bibitem {Rakhmanova} {S. Rakhmanova and D. L. Mills, Phys. Rev. B {\bf 58}, 11458 (1998).}

\bibitem {Slavin} {A. N. Slavin, Yu. S. Kivshar, E. A. Ostrovskaya, and H. Benner, Phys.
Rev. Lett. {\bf 82}, 2583 (1999).}

\bibitem {Bischof} {B. Bischof, A. N. Slavin, H. Benner, and Yu. S. Kivshar, Phys. Rev. B
{\bf 71}, 104424 (2005).}

\bibitem {Zaspel} {C. E. Zaspel, J. H. Mantha, Yu. G. Rapoport, and V. V. Grimalsky, Phys.
Rev. B {\bf 64}, 064416 (2001).}

\bibitem {Ivanov} {B. A. Ivanov, A. Yu. Galkin, R. S. Khymyn, and A. Yu. Merkulov, Phys.
Rev. B {\bf 77}, 064402 (2008).}

\bibitem {Kivshar98} {Yu. S. Kivshar and B. Luther-Davies, Phys. Reports {\bf
298}, 81 (1998).}

\bibitem {Kivshar94} {Yu. S. Kivshar, W. Kr\'{o}likowski, and A. O. Chubykalo, Phys. Rev. E
{\bf 50}, 5020 (1994).}

\bibitem {Kevrekidis} {P. G. Kevrekidis, I. G. Kevrekidis, A. R. Bishop, and E. S. Titi, Phys.
Rev. E {\bf 65}, 046613 (2002).}

\bibitem {Forinash} {K. Forinash, M. Peyrard, and B. Malomed, Phys. Rev. E {\bf
49}, 3400 (1994).}

\bibitem {Burtsev} {S. Burtsev, D. J. Kaup, and B. A. Malomed, Phys. Rev. E {\bf
52}, 4474 (1995).}

\bibitem {Sukhorukov} {A. A. Sukhorukov, Yu. S. Kivshar, O. Bang, J. J. Rasmussen,
and P. L. Christiansen, Phys. Rev. E {\bf 63}, 036601 (2001).}

\bibitem {Frantz} {D. J. Frantzeskakis, G. Theocharis, F. K. Diakonos, P. Schmelcher,
and Yu. S. Kivshar, Phys. Rev. A {\bf 66}, 053608 (2002).}

\bibitem {Prima2009} {M. T. Primatarowa, R. S. Kamburova, and K. T. Stoychev,
J. Optoel. Adv. Mat. {\bf 11}, 1388 (2009).}

\bibitem {Lai} {R. Lai, S. A. Kiselev, and A. J. Sievers, Phys. Rev.
B {\bf 56}, 5345 (1997).}

\bibitem {Prima2002} {M. T. Primatarowa, K. T. Stoychev, and R. S. Kamburova, Eur. Phys. J.
B {\bf 29}, 291 (2002).}

\bibitem {Prima95} {M. T. Primatarowa, K. T. Stoychev, and R. S. Kamburova, Phys. Rev.
B {\bf 52}, 15291 (1995).}

\end{thebibliography}
\end{document}